\documentclass{jpsj3}

\newcommand{\subcaption}[1]{\begin{center}#1\end{center}}

\makeatletter
\newcommand\newblock{\hskip .11em\@plus.33em\@minus.07em}
\makeatother

\usepackage{txfonts}
\usepackage{authblk}
\usepackage{url}

\usepackage[utf8]{inputenc}
\usepackage[T1]{fontenc}
\usepackage{qcircuit}
\usepackage{braket}

\usepackage{graphicx}
\graphicspath{ {images/} }

\usepackage{bm}

\title{Storage Capacity Evaluation of the Quantum Perceptron using the Replica Method}
\author[1,2]{Mitsuru Urushibata}
\author[1,3,4]{Masayuki Ohzeki}

\affil[1]{Graduate School of Information Sciences, Tohoku University, Sendai 980-8579, Japan}
\affil[2]{Crosstab Inc., Kawasaki 211-0068, Japan}
\affil[3]{Sigma-i Co., Ltd., Tokyo 108-0075, Japan}
\affil[4]{Department of Physics, Tokyo Institute of Technology, Meguro, 152-8551 Japan}

\abst{We investigate a quantum perceptron implemented on a quantum circuit using a repeat until method. 
We evaluate this from the perspective of capacity, one of the performance evaluation measures for perceptions.
We assess a Gardner volume, defined as a volume of coefficients of the perceptron that can correctly classify given training examples using the replica method. 
The model is defined on the quantum circuit.
Nevertheless, it is straightforward to assess the capacity using the replica method, which is a standard method in classical statistical mechanics.
The reason why we can solve our model by the replica method is the repeat until method, in which we focus on the output of the measurements of the quantum circuit.
We find that the capacity of a quantum perceptron is larger than that of a classical perceptron since the quantum one is simple but effectively falls into a highly nonlinear form of the activation function.
}

\inst{}

\begin{document}
\maketitle

\section{Introduction}
Recently, deep learning has been studied and applied in various fields. For example, it has important applications in areas such as image and speech recognition\cite{krizhevsky2012imagenet,soltau2016neural}, and particularly in generative AI\cite{goodfellow2014generative,wang2021generative,vaswani2017attention}. 
Deep-learning-based algorithms are composed of a multi-layer neural network; increasing the number of layers can potentially achieve higher learning and representational capabilities. 
Perceptrons are the basic components of neural networks, they are simple supervised machine-learning models proposed by Rosenblatt\cite{rosenblatt1958perceptron} that employ the Hebbian rule.
Analysis of the perceptron model is suitable as the first step of the non-trivial aspects of a neural network with data and is essential for advancing deep learning \cite{Ohzeki2015,Arai2021,Gerace2024}. 
A perceptron can identify linearly separable data. 
However, it cannot identify cases similar to those involving an XOR gate. 
Therefore, it is important to understand the limitations of perceptrons in terms of their discriminative ability. 
Cover \cite{cover1965geometrical,orhan2014cover} demonstrated the relationship between the input data dimension $N$, number of data instances to be classified $P$, and number of separable input patterns achievable with perception. 
Cover's theorem states that for a sufficiently large $N$, when $P/N$ $> 2$, the discriminative ability diminishes and there are no cases where a correct classification is possible. 
In other words, it indicates a critical point in the relationship between the input dimension and number of instances. 
This theorem is proven using a combinatorial mathematical method. Gardner \cite{gardner1988optimal,E.Gardner_1987} proposed another approach using the spin glass theory and calculated the volume in the perceptron coefficient space, which completely identified all patterns. Assuming that the data were randomly provided, the configurational average of this volume was calculated using the replica method, leading to the evaluation of the saddle point equations and resulting in the critical point $\alpha_{c}=2$. It was recently found that increasing the hidden layer depth in deep learning results in networks with larger capacity\cite{yoshino2020complex,PhysRevResearch.5.033068}. Several related studies have been conducted \cite{shinzato2008perceptron,krauth1989storage,boffetta1993symmetry,crespi1999storage,nokura1994capacity,zavatone2021activation}

We considered extending this capacity estimation problem to a quantum version of the perceptron.
A comparison between quantum and classical perceptrons is interesting from a storage capacity perspective. 
Aikaterini et al. \cite{tacchino2019artificial} computed the storage capacity of a previously proposed perceptron by using the replica method\cite{kasper2021storage}. 
They demonstrated that the storage capacity of this quantum perceptron for spherical weights was greater than that for classical perception. They stated that this is because the perceptron has a quadratic form rather than a linear form. In a similar study\cite{benatti2022pattern}, the capacity of continuous variable quantum perceptrons\cite{benatti2019continuous} was calculated, demonstrating that their quantum perceptron did not surpass classical perceptrons.

Among the various quantum perceptron algorithms\cite{lewenstein1994quantum,kapoor2016quantum,cao2017quantum,tacchino2019artificial}, this study focused on Cao’s method\cite{cao2017quantum} which uses quantum gates called repeat-until-success (RUS) circuits.
Using this approach, non-linear effects were introduced into a quantum perceptron. Using Gardner’s method, we demonstrate that Cao's quantum perceptron has a higher storage capacity than the classical perceptron. 
The remainder of this paper is organized as follows. In Section 2, we first introduce the basic settings regarding Cao's perception, storage capacity, Gardner volume, and the replica method. 
In Section 3, we describe
and evaluate the saddle point equation, obtaining the corresponding storage capacity. 
Finally, in Section 4, we summarize our results and discuss future research prospects. Quantum machine learning is expected to offer advantages such as speed and improved accuracy; however, whether it is superior to classical methods remains unclear. We believe that this study addresses this question.

\section{Problem settings}
\subsection{Quantum perceptrons}
\begin{figure}[htbp]
\begin{tabular}{c}
      \begin{minipage}{0.5\hsize}
\centering
        \[
        \Qcircuit @C=1.0em @R=0.7em @!R {
            & \lstick{\ket{x}} & \ctrl{1} &  \qw  & \ctrl{1} & \qw &  \\
            & \lstick{\ket{0}} & \gate{R_{y}(2\theta)} & \ctrl{1} & \gate{R_{y}^{\dagger}(2\theta)} & \meter  \\
            & \lstick{\ket{0}} & \qw & \gate{-iY} & \qw & \rstick{\qquad \raisebox{3.5em}{$\alpha \ket{0} \otimes R_{y}(2\phi_{1}(\theta))\ket{0} + \beta \ket{1} \otimes R_{y}(\pi/2) \ket{0}$}} \qw 
            \gategroup{2}{3}{3}{6}{2.0em}{\}}
            }
        \]
    \subcaption{(a)}
      \end{minipage} \\
      \begin{minipage}{0.5\hsize}
\centering
        \[
        \Qcircuit @C=1.0em @R=0.7em @!R {
            & \lstick{\ket{x}} & \ctrl{1} &  \qw  & \ctrl{1} & \qw &  \qw \\            & \lstick{\ket{0}} & \gate{R_{y}(2\phi_{k-1}(\theta))} & \ctrl{1} & \gate{R_{y}^{\dagger}(2\phi_{k-1}(\theta))} & \qw  &\rstick{\ket{0}}  \qw  \\
            & \lstick{\ket{0}} & \qw & \gate{-iY} & \qw & \qw&\rstick{R_{y}(2\phi_{k}(\theta))\ket{0}}  \qw 
}
        \]
        \subcaption{(b)}
      \end{minipage}
      \end{tabular}
    \caption{(a) Process in the initial step, where output $R(\phi_{1}(\theta))$ is obtained from input $\theta = \bm{w}\bm{x}$. When the outcome is $0$, the operation is repeated with $\phi_{1}(\theta)$ as the new $\theta$. When the outcome is $1$, the process starts over and repeats the same operation. (b) Circuit after repeating the operation k times. Here, $\phi_{k}(\theta)$ is defined as $\arctan(\tan^{2^k}(\theta))$.}
    \label{fig:gate1}
\end{figure}
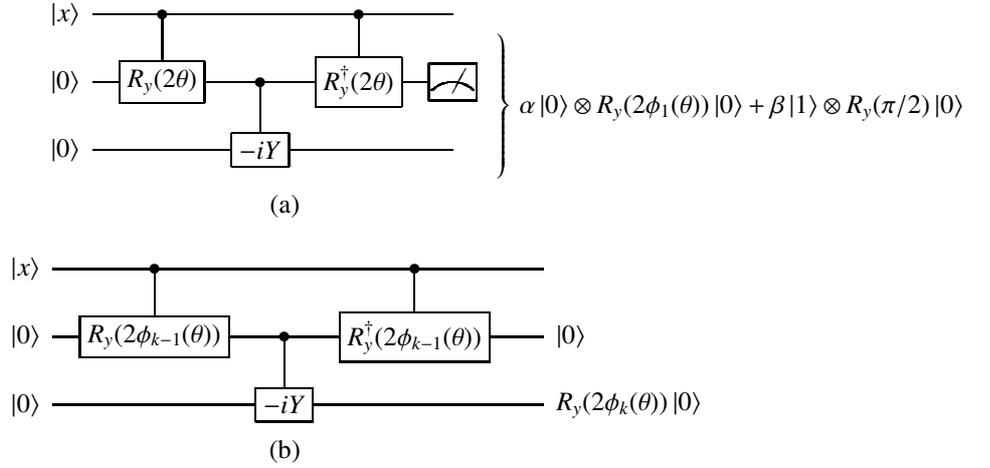

The quantum perceptron proposed by Cao et al. \cite{cao2017quantum} is defined by a quantum circuit and maps $N$-dimensional input data $\bm{x} \in \{-1,1\}^N$ to label $y \in \{-1,1\}$, where $-1$ and $1$ correspond to quantum states $\ket{0}$ and $\ket{1}$, respectively. 
Let $(\bm{x},y)$ be a pair of input data and labels and $\bm{w}=(w_{1},w_{2},...,w_{N})^\top$ be the perceptron weights. 
When the $i$th element of the input data $x_{i}$ equals $1$, the target bit rotates by $2w_{i}$ along the $y$-axis. 
Applying this operation to $1\leq i \leq N$, the state vector is acted upon by $R_{y}(2\theta)$, where $\theta=\bm{w}^\top\bm{x}$. 
The quantum circuit is shown in Fig. \ref{fig:gate1}. 
We measured the ancillary qubit, and when the outcome was $0$, the operation was repeated with $\arctan(\tan^{2}(\theta))$ as the new $\theta$. For an outcome of $1$, the process starts over and repeats the same operation. Such a circuit is referred to as a repeat-until-success circuit \cite{wiebe2014quantum}. This is repeated $k$ times and the output register state becomes $R_{y}(2\arctan(\tan^{2^{k}}(\theta))) \ket{0}$. 
The relationship between the output and input is as follows:
\begin{eqnarray}
    P\left(1|\bm{w},\bm{x}\right) = \bra{0}R_{y}\left(2\arctan\left(\tan^{2^{k}}(\bm{w}^\top\bm{x})\right)\right)^{\dagger} \ket{1}\bra{1}R_{y}\left(2\arctan\left(\tan^{2^{k}}\left(\bm{w}^\top\bm{x}\right)\right)\right) \ket{0} .
\end{eqnarray}
When $\bm{w}^{t} \bm{x}$ is close to $0$, the output $-1$ is more likely to be observed; when it is close to $\pi/4$, $1$ is more likely to be observed as the output. 
This exhibits a non-linear effect similar to that of the sigmoid function. 
In an actual quantum device, measurements are obtained based on this distribution. 
We define the output value as the expected value.
\begin{eqnarray}
\label{eq:cao}
    \hat{y}\left(\bm{w},\bm{x}\right) = -\bra{0}R_{y}\left(2\arctan\left(\tan^{2^{k}}\left(\bm{w}^\top\bm{x}\right)\right)\right)^{\dagger} Z R_{y}\left(2\arctan\left(\tan^{2^{k}}\left(\bm{w}^\top\bm{x}\right)\right)\right) \ket{0},
\end{eqnarray}
where $Z$ denotes the Pauli $Z$-operator. 
When $y > 0$, it is classified as $1$; otherwise, it is classified as $-1$. If the label is correctly classified, $\hat{y}y > 0$ holds true. 
By converting the input data into binary form, the perceptron can be extended to general cases, such as real-number inputs. 
This solution can be obtained by applying the perceptron learning algorithm to the above output on a classical computer. This learning method is called a hybrid quantum-classical algorithm.
\subsection{Storage Capacity}
Let $N$ be the input data dimension and $p$ be the number of input data points. 
The number of linearly separable label patterns is $C(p,N)$. 
For $p_{N} \equiv \max \left\{p \colon {C(p,N)}/{2^p}=1 \right\}$, the perception storage capacity is defined as follows\cite{cruciani2022capacity}:
\begin{eqnarray}
    \alpha_{c} = \lim_{N \to \infty} \frac{p_{N}}{N}.
\end{eqnarray}
For a classical binary perceptron, $\alpha_{c}$ is $2$.
\subsection{Gardner Volume and Replica Method}
In general, for $(\bm{x},y) \in \mathbb{R}^{N} \times \{-1,1\}, \bm{w} \in \mathbb{R}$, the perceptron output is defined as follows:
\begin{eqnarray}
\hat{y}\left(\bm{w},\bm{x}\right) = \left\{
\begin{array}{ll}
1 & (\bm{w}^\top \bm{x} \geq 0) \\
-1 & (\text{otherwise}).
\end{array}
\right.
\end{eqnarray}
Gardner introduced the space volume of coefficients $\bm{w}$, which completely identifies all patterns; this is called the Gardner Volume\cite{gardner1988optimal,gardner1987maximum}. 
Let $\{(\bm{x_{\mu}},y_{\mu})\}_{\mu=1}^{P}$ be P pairs of input data and labels given randomly, and $\bm{w}^\top \bm{w}=N$ be constraints, then the volume can be written as
\begin{eqnarray}
    V = \int d\bm{w} \delta\left(\bm{w}^\top \bm{w}-N\right) \prod_{\mu=1}^{P} \Theta\left(y_{\mu}\hat{y}_{\mu}\left(\bm{w},\bm{x}_{\mu}\right)\right) ,
\end{eqnarray}
where $\Theta$ denotes the Heaviside step function. 
To evaluate the typical property of $V$, we calculate the average of $\log{V}$ on the $\bm{x}$ and $y$ distributions. 
In general, it is difficult to calculate this value directly.
Accordingly, we employed the replica trick, which is described as follows:
\begin{eqnarray}
    \left[ \log{V} \right] = \lim_{n \to 0} \frac{[V^n]-1}{n},
\end{eqnarray}
\begin{eqnarray}
    [V^n] = \int \prod_{a=1}^{n} d\bm{w}_{a} \delta\left(\bm{w}_{a}^{\top} \bm{w}_{a}-N\right) \prod_{\mu=1}^{P}\left[\prod_{a=1}^{n}\Theta\left(y_{\mu}\hat{y}_{\mu}\left(\bm{w}_{a},\bm{x}_{\mu}\right)\right)\right], 
\end{eqnarray}
where $[...]$ denotes the average on $p(\bm{x}),p(y)$. Furthermore, by setting the overlap between the replicas to $\bm{w}_{a}^\top\bm{w}_{b}/N=q_{ab}$, $[V^n]$ is expressed as 
\begin{eqnarray}
    [V^n] = \int dQ \prod_{a=1}^{n} d\bm{w}_{a} \prod_{a,b}^{n}\delta\left(\bm{w}_{a}^{\top} \bm{w}_{b}-q_{ab}N\right) \prod_{\mu=1}^{P}\left[\prod_{a=1}^{n}\Theta\left(y_{\mu}\hat{y}_{\mu}\left(\bm{w}_{a},\bm{x}_{\mu}\right)\right)\right],
    \label{eq:vn}
\end{eqnarray}
where $Q \equiv \left(q_{ab}\right) $. 
The energy and entropy terms are formally defined as
\begin{eqnarray}
    \label{eq:terms}
 \exp\left(-N \beta e(Q) \right) = \prod_{\mu=1}^{P}\left[\prod_{a=1}^{n}\Theta\left(y_{\mu}\hat{y}_{\mu}\left(\bm{w}_{a},\bm{x}_{\mu}\right)\right)\right], \\
 \exp\left(NS(Q) \right) = \prod_{a=1}^{n} \int  d\bm{w}_{a} \prod_{a,b}^{n}\delta\left(\bm{w}_{a}^{\top} \bm{w}_{b}-q_{ab}N\right), 
\end{eqnarray}
Then, we can rewrite (\ref{eq:vn}) as
\begin{eqnarray}
[V^n] &=& \int dQ \exp\left(N\left(-\beta 
 e(Q)+S(Q)\right)\right) \\
&\approx& \exp\left(N\left(-\beta e(\hat{Q})+S(\hat{Q})\right)\right) \quad\text{when $N>>1$},
\end{eqnarray}
where $\hat{Q}$ is a saddle point and $\beta$ is the inverse temperature.
Here, we assume that $\{\bm{x}_{\mu}\}_{\mu=1}^{p}$ is drawn i.i.d. from a Gaussian distribution $\bm{x}_{\mu} \sim N(0,1/N)$ and $y_{\mu}$ follows a binomial distribution. 
Finally, We assume the following replica symmetry for the relationship between replicas.
\begin{eqnarray}
q_{ab} = 
\left\{
  \begin{array}{ll}
    q, & \text{if } a \neq b \\
    1, & \text{if } a=b
  \end{array}
\right.
\end{eqnarray}
\section{Results}
\begin{figure}[htbp]
    \centering
    \includegraphics[width=15.0cm]{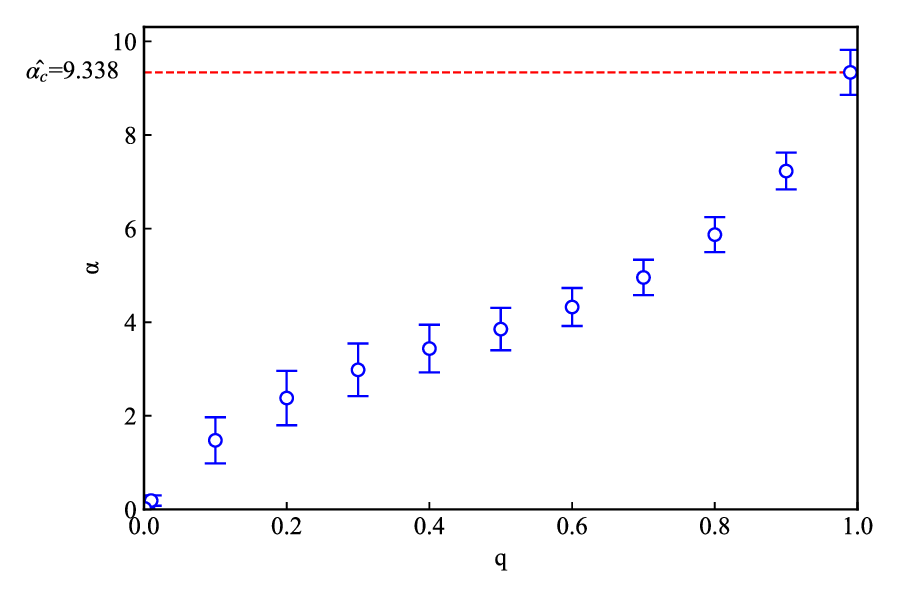}
    \caption{(Color) $q$ results. The horizontal axis represents the order parameter, while the vertical axis represents the estimates of the equation for the corresponding $q$ value. As $q$ approaches $1$, $\alpha$ approaches the capacity. The estimated value is approximately $9.338$, which is larger than that of the well-known classical perceptron. The error bars represent the standard errors of the estimates.}
    \label{fig:result}
\end{figure}

We evaluated the volume of Cao's quantum perceptron. 
First, we transform (\ref{eq:cao}) into a more straightforward expression, as follows:
\begin{eqnarray}
    \hat{y}\left(\bm{w},\bm{x}_{\mu}\right)  &=& -\left(\cos^2\left(\arctan\left(\tan^{2^k}\left(\bm{w}^\top\bm{x}_{\mu}\right)\right)\right) - \sin^2\left(\arctan\left(\tan^{2^k}\left(\bm{w}^\top\bm{x}_{\mu}\right)\right)\right)\right) \\
    &=& -\cos\left(2\arctan\left(\tan^{2^k}\left(\bm{w}^\top\bm{x}_{\mu}\right)\right)\right).
\end{eqnarray}
We apply the above equation to (\ref{eq:vn}) and focus on the Heaviside step function:
\begin{eqnarray}
    \label{eq:mean}
    \left[\prod_{a=1}^{n}\Theta\left(-y_{\mu}\cos\left(2\arctan\left(\tan^{2^k}\left(\bm{w}_{a}^\top\bm{x}_{\mu}\right)\right)\right)\right)\right] = \prod_{a=1}^{n}\int_{D}d\bm{x}_{\mu}d y_{\mu} p (\bm{x}_{\mu}) p(y_{\mu}),
\end{eqnarray}
where $D=\bigcup_{l \in \mathbb{Z}} \left\{(\bm{x_{\mu}},y_{\mu}) \mid l\pi < y_{\mu}\left(\bm{w}_{a}^{\top} \bm{x}_{\mu}-\pi/4 \right) < l\pi + \pi/2 \right\}$, because $\Theta(...)=1$ is equivalent to $(\bm{x_{\mu}},y_{\mu}) \ \in D$. 
The derivation of $D$ is presented in the appendix. 
As $y_{\mu}(\bm{w}_{a}^{\top} \bm{x}_{\mu}-\pi/4)$ follows $N(0,1+\pi^2/16)$, this can be expressed as 
\begin{eqnarray}
y_{\mu}\left(\bm{w}_{a}^{\top} \bm{x}_{\mu}-\frac{\pi}{4}\right)  = \sqrt{q+\frac{\pi^2}{16}}z+\sqrt{1-q}X_{a},
\end{eqnarray}
where $z\sim N(0,1)$, $X_{a} \sim N(0,1)$. 
Following the above variable transformation, (\ref{eq:mean}) can be rewritten as follows:
\begin{eqnarray}
    \prod_{a=1}^{n}\int_{D}Dz DX_{a} &=& \int Dz  \left(\int_{I} DX \right)^n \\
    &=& \int Dz \exp \left(n \log \left(\int_{I} DX \right)\right) \\
    & \approx & 1 + n \int Dz \log \left(\int_{I} DX \right),
    \label{eq:energy}
\end{eqnarray}
where $I=\bigcup_{l \in \mathbb{Z}} \left[\left(-\sqrt{q+\pi^2/16}z+l\pi\right)/\sqrt{1-q},\left(-\sqrt{q+\pi^2/16}z+l\pi+\pi/2\right)/\sqrt{1-q} \right]$. 
By substituting (\ref{eq:energy}) into (\ref{eq:terms}), we obtain
\begin{eqnarray}
-\beta e(Q) & \approx & \frac{1}{N} \sum_{\mu=1}^{P} \log \left(1 + n \int Dz \log \left(\int_{I} DX \right)\right) \\
\label{eq:eq}
& \approx & n\frac{P}{N} \int Dz \log \left(\int_{I} DX \right).
\end{eqnarray}
Next, we assess $S(Q)$. 
Employing the Fourier representation of Dirac's delta function, we obtain the following:
\begin{eqnarray}
\exp(NS(Q)) =  \prod_{a=1}^{n} \int d \bm{w}_{a}\prod_{a \neq b} \int d\tilde{q}\exp\left(-\frac{\tilde{q}}{2} (Nq-\bm{w_{a}^{t}}\bm{w_{b}}) \right) \prod_{a=1}^n \int d\tilde{t}\exp\left(\frac{\tilde{t}}{2} (N-\bm{w_{a}^{t}}\bm{w_{a}}) \right) \\
\label{eq:nsq}
\approx \int d\tilde{q} \int d\tilde{t}\exp \left\{ nN\left(-\frac{\log(\tilde{t} + \tilde{q})}{2}+ \frac{\left(q \tilde{q} + \tilde{t}\right)}{2}  +\frac{\tilde{q}}{2(\tilde t +\tilde q)} \right)\right\}C, 
\end{eqnarray}
where $C$ is constant. Further details on this calculation can be found in the appendix. Summarizing (\ref{eq:eq}) and (\ref{eq:nsq}), we derive the following equation:
\begin{eqnarray}
[V^n] \approx \int dq  \exp\left(nN\alpha \int Dz \log \left(\int_{I} DX \right)\right) \int d\tilde{q} d\tilde{t}\exp \left\{ nN\left(-\frac{\log(\tilde{t} + \tilde{q})}{2}+ \frac{\left(q \tilde{q} + \tilde{t}\right)}{2}  +\frac{\tilde{q}}{2(\tilde t +\tilde q)} \right)\right\}C.
\end{eqnarray}
When we differentiate $q$, $\tilde q$, and $\tilde t$ and solve for the saddle point, we obtain the following saddle-point equation:
\begin{eqnarray}
\alpha = \frac{-q}{2(1-q)^2} \left(\int Dz \frac{\partial}{\partial q} \log \left(\int_{I} DX \right) \right)^{-1}, 
\end{eqnarray}
\begin{eqnarray}
\frac{\partial}{\partial q} \log \left(\int_{I} DX \right) = 
\frac{\sum_{l \in \mathbb{Z}}\left( \frac{\partial}{\partial q}\left( G_{1}(q,z,l)\right) f\left(G_{1}(q,z,l)\right) - \frac{\partial}{\partial q} \left(G_{2}(q,z,l)\right) f\left(G_{2}(q,z,l)\right)\right)}{\sum_{l \in \mathbb{Z}}\left(F\left(G_{1}(q,z,l)\right) - F\left(G_{2}(q,z,l)\right)\right)}, 
\label{eq:derive}
\end{eqnarray}
where
\begin{eqnarray}
G_{1}(q,z,l) = \frac{-\sqrt{q+\pi^2/16}z+l\pi  +\pi/2 }{\sqrt{1-q}}, \quad G_{2}(q,z,l) = \frac{-\sqrt{q+\pi^2/16}z +l\pi  }{\sqrt{1-q}}.
\end{eqnarray}
Function $f$ in (\ref{eq:derive}) is the probability density function of the standard normal distribution and $F$ is the cumulative distribution function. 
We evaluated the above equation numerically. 
When $q$ approaches $1$, we obtain the critical point for the storage capacity. The data shown in Fig. \ref{fig:result} suggests the possibility that it exceeds the classical perception capacity under the assumption of replica symmetry.
\section{Conclusion and Future work}
This study was inspired by Cao et al. \cite{cao2017quantum}, who introduced a perceptron using the repeat-until-success (RUS) method on a quantum circuit. 
In this study, we extend their perceptron to handle real-valued inputs and investigated them from a capacity perspective. 
Based on the results, we demonstrated that the proposed perceptron surpasses the capacity of classical perception under the assumption of replica symmetry. 
We believe that this is because the perceptron outputs exhibit periodicity for the inputs. 
Because this property involves multiple linear separation planes, the perceptron works well even for complex data distributions. 
However, this result is not due to quantum effects such as entanglement or superposition in the perceptron; instead, the algebraic contributions of the quantum circuit play a significant role. 
Activation functions with periodicity are commonly believed to be difficult to train\cite{parascandolo2016taming,sitzmann2020implicit}. 
Learning real data and validating them remain tasks yet to be accomplished.

There are two major directions for future research. 
The first is to examine the stability of the replica-symmetric solutions. 
In the case of non-monotonic perceptrons, replica-symmetric solutions are known to become unstable, and it has been shown that one-step replica-symmetric-broken ($1$-RSB) solutions have a smaller storage capacity than replica-symmetric (RS) solutions\cite{boffetta1993symmetry,nishimori2001statistical}. 
Because the perceptron considered in this study is also non-monotonic, further solution stability verification is required.
The second is to investigate the generalization of the proposed perceptron using a supervised replica. 
We assume an appropriate teacher machine and calculate its overlap with the student machine. 
According to previous studies, the generalization error is given as a function of the overlap between the teacher and students\cite{T.L.H.Watkin_1993}. 
As an extension of this study, it would be interesting to evaluate the generalization error of this quantum perceptron. 
We believe that these potential future studies will contribute to the development of quantum machine learning and statistical mechanics.

\begin{acknowledgment}
This work is supported by JSPS KAKENHI Grant No. 23H01432.
Our study receives financial support from the programs for Bridging the gap between R\&D and the IDeal society (society 5.0) and Generating Economic and social value (BRIDGE) and the Cross-ministerial Strategic Innovation Promotion Program (SIP) from the Cabinet Office.
\end{acknowledgment}

\newpage
\bibliographystyle{jpsj}
\bibliography{mybibliography.bib}

\newpage
\appendix
\section{}
\subsection{Derivation of the set of integration in the equation (\ref{eq:mean})}
if $y_{\mu}=+1$ then,
\begin{eqnarray*}
\theta \left(-\cos\left(2\arctan\left(\tan^{2^k}\left(\bm{w}^\top\bm{x}\right)\right)\right)\right) =1 \\
 \iff  -\cos\left(2\arctan\left(\tan^{2^k}\left(\bm{w}^\top\bm{x}\right)\right)\right)>0 \\
 \iff  \frac{\pi}{4} < \arctan\left(\tan^{2^k}\left(\bm{w}^\top\bm{x}\right)\right) < \frac{\pi}{2} \\
\iff 1< \tan^{2^k}\left(\bm{w}^\top\bm{x}\right) < \infty \\
\iff l \pi < \bm{w}^\top\bm{x} - \frac{\pi}{4} < l \pi + \frac{\pi}{2} \quad (l \in \mathbb{Z}).
\end{eqnarray*}
Inversely, setting $y_{\mu} = -1$ we get the following result,
\begin{eqnarray*}
-l \pi < (-1)\left(\bm{w}^\top\bm{x} - \frac{\pi}{4}\right) < -l \pi + \frac{\pi}{2} \quad (l \in \mathbb{Z}).
\end{eqnarray*}
\subsection{Derivation of the equation (\ref{eq:nsq})}
Employing Fourier Representation of Dirac's Delta Function, we can transform (\ref{eq:nsq}) into the following expression,
\begin{eqnarray*}
\exp(NS(Q)) &=& \prod_{a=1}^{n} \int d \bm{w}_{a} \prod_{a \neq b}\delta(\bm{w_{a}^{\top}}\bm{w_{b}}-Nq) \prod_{a=1}^n \delta(\bm{w_{a}^{\top}}\bm{w_{a}}-N) \\
&=&  \prod_{a=1}^{n} \int d \bm{w}_{a}\prod_{a \neq b} \int d\tilde{q} \exp\left({-\frac{\tilde{q}}{2} (Nq-\bm{w_{a}^\top}\bm{w_{b}})}\right)  \prod_{a=1}^n \int d\tilde{t} \exp \left({\frac{\tilde{t}}{2} (N-\bm{w_{a}^{\top}}\bm{w_{a}}) }\right) \\
&=&  \prod_{a=1}^{n} \int d \bm{w}_{a} \int d\tilde{q} \int d\tilde{t} \exp \left({-\frac{\tilde{q}}{2} \sum_{a\neq b}(Nq-\bm{w_{a}^{\top}}\bm{w_{b}}) } \right)  \exp \left({\frac{\tilde{t}}{2} \sum_{a=1}^n(N-\bm{w_{a}^{\top}}\bm{w_{a}}) }\right).
\end{eqnarray*}
The argument of the above exponential function is,  
\begin{eqnarray*}
 -\frac{{N}n(n-1)}{2}\tilde{q}q+\frac{nN}{2}\tilde{t} + \frac{\tilde{q}}{2}\sum_{a\neq b} \bm w_{a}^t \bm w_{b} -\frac{\tilde{t}}{2} \sum_{a=1}^n \bm w_{a}^t \bm w_{a} \\
\thickapprox \frac{nN}{2}\left(q \tilde{q} + \tilde{t}\right) + \frac{\tilde{q}}{2} \left\{ \left(\sum_{a=1}^n \bm w_{a}\right)^2 - \sum_{a=1} \bm w_{a}^2\right\} -\frac{\tilde{t}}{2} \sum_{a=1}^n \bm w_{a}^t \bm w_{a} 
\end{eqnarray*}
The above approximation arises from $n^2 \approx 0$ $(n \ll 1)$. Applying Gaussian integration to the second term, 
\begin{eqnarray*}
\exp \left( \frac{\tilde{q}}{2}  \left(\sum_{a=1}^n \bm w_{a}\right)^2 \right) = \int D {Z} \exp(\sqrt{q} z\sum_{a=1}^n \bm w_{a}).
\end{eqnarray*}
We can continue to calculate the $\exp(NS(Q))$, 
\begin{eqnarray*}
\exp(NS(Q)) &=& \prod_{a=1}^{n} \int d \bm{w}_{a} \int d\tilde{q} \int d\tilde{t}\int DZ\exp \left(\frac{nN}{2}\left(q \tilde{q} + \tilde{t}\right) - \frac{1}{2}(\tilde t +\tilde q)\sum_{a=1}^n \bm w_{a}^t \bm w_{a} + \sqrt{q} z\sum_{a=1}^n \bm w_{a} \right)  \\
&=&  \int d\tilde{q} \int d\tilde{t}\int D Z \prod_{a=1}^n\int d \bm{w}_{a} \exp \left(\frac{nN}{2}\left(q \tilde{q} + \tilde{t}\right) - \frac{1}{2}(\tilde t +\tilde q) (\bm w_{a}-  \frac{\sqrt{\tilde{q}}}{(\tilde t +\tilde q)} z)^2 +\frac{\tilde{q}z^2}{2(\tilde t +\tilde q)} \right)  \\
&=& \int d\tilde{q} \int d\tilde{t}\int DZ \left(\sqrt{\frac{2\pi}{\tilde{t} + \tilde{q}}}\right)^{nN}\exp \left(\frac{nN}{2}\left(q \tilde{q} + \tilde{t}\right)  +nN\frac{\tilde{q} z^2}{2(\tilde t +\tilde q)} \right)  \\
&=& \int d\tilde{q} \int d\tilde{t} \int DZ \exp \left(\frac{nN}{2}\log(2\pi) -\frac{nN}{2}\log(\tilde{t} + \tilde{q})+ \frac{nN}{2}\left(q \tilde{q} + \tilde{t}\right)  +nN\frac{\tilde{q}z^2}{2(\tilde t +\tilde q)} \right)  \\
& \approx & \int d\tilde{q} \int d\tilde{t}  \exp \left(\frac{nN}{2}\log(2\pi) -\frac{nN}{2}\log(\tilde{t} + \tilde{q})+ \frac{nN}{2}\left(q \tilde{q} + \tilde{t}\right)  \right) \left(\int DZ \left(1 + nN\frac{\tilde{q}z^2}{2(\tilde t +\tilde q)} \right)\right)   \\
& = & \int d\tilde{q} \int d\tilde{t}  \exp \left(\frac{nN}{2}\log(2\pi) -\frac{nN}{2}\log(\tilde{t} + \tilde{q})+ \frac{nN}{2}\left(q \tilde{q} + \tilde{t}\right)  \right)  \left(1 + nN\frac{\tilde{q}}{2(\tilde t +\tilde q)} \right)   \\
& \approx & \int d\tilde{q} \int d\tilde{t}\exp \left\{ nN\left(-\frac{\log(\tilde{t} + \tilde{q})}{2}+ \frac{\left(q \tilde{q} + \tilde{t}\right)}{2}  +\frac{\tilde{q}}{2(\tilde t +\tilde q)} \right)\right\}C,  
\end{eqnarray*}
where $DZ$ is  $dz (1/\sqrt{2\pi}) \exp(-z^2/2)$ and $C$ is constant.
\end{document}